%% file: main.tex
\documentclass{article}
\usepackage{spconf,amsmath,graphicx}
\usepackage{booktabs}
\usepackage{amsfonts}
\usepackage{multirow}
\usepackage{bbding}
\usepackage{cite}
\usepackage[urlcolor=blue]{hyperref}
\usepackage{subfigure}
\usepackage{mathrsfs}
\usepackage{threeparttable}
\usepackage{url}
\title{The THU-HCSI multi-speaker multi-lingual few-shot voice cloning system for LIMMITS'24 Challenge}

\name{Yixuan Zhou$^{1}$, Shuoyi Zhou$^{1}$, Shun Lei$^{1}$, Zhiyong Wu$^{1*}$\thanks{$^{*}$ Corresponding author.}, Menglin Wu$^{2}$}
\address{
  $^1$Shenzhen International Graduate School, Tsinghua University, Shenzhen, China\\
  $^2$ByteDance, Shanghai, China \\
    \small{
        \{yx-zhou23, zhousy23, leis21\}$@$mails.tsinghua.edu.cn, zywu$@$sz.tsinghua.edu.cn,
        menglinwoo@gmail.com} \\ 
}

\begin{document}
\ninept
\maketitle
\input{Paper_Parts/Abstract}
\input{Paper_Parts/Introduction}
\input{Paper_Parts/Methodology}

\input{Paper_Parts/Results}

\input{Paper_Parts/Conclusions}

\bibliographystyle{IEEEbib}
\bibliography{main}

\end{document}

%% file: Paper_Parts/Abstract.tex
\begin{abstract}
This paper presents the multi-speaker multi-lingual few-shot voice cloning system developed by THU-HCSI team for LIMMITS'24 Challenge.
To achieve high speaker similarity and naturalness in both mono-lingual and cross-lingual scenarios, we build the system upon YourTTS and 
add several enhancements.
For further improving
speaker similarity and speech quality, we introduce speaker-aware text encoder and flow-based decoder with Transformer blocks.
In addition, we denoise the few-shot data, mix up them with pre-training data, and adopt a speaker-balanced sampling strategy to guarantee effective fine-tuning for target speakers.
The official evaluations in track 1 show that  our system achieves the best
speaker similarity MOS of 4.25 and obtains considerable naturalness MOS of 3.97.

\end{abstract}
\vspace{-0.1cm}

%% file: Paper_Parts/Introduction.tex
\vspace{-0.3cm}
\section{Introduction}
\label{sec:intro}
\vspace{-0.15cm}

Text-to-speech (TTS) models have made great progress in recent years, and TTS research for low-resource scenarios, such as some Indian languages, has become popular and necessary.
The LIMMITS'24, as a part of ICASSP Signal Processing Grand Challenge, aims to provide the opportunity for the participants to perform TTS voice cloning with an Indian multilingual base model \cite{limmit24}.
In track 1, participants can pre-train the base model with provided challenge dataset, and perform few-shot voice cloning by fine-tuning on new speakers' data. 
In track 2, external datasets are allowed for pre-training, while track 3 focuses on zero-shot scenarios. 
The evaluations are conducted on both mono-lingual and cross-lingual synthesis, with naturalness and speaker similarity subjective tests. 

In this work, we develop a few-shot TTS voice cloning system to support multiple speakers and languages, and submit it to track 1.
To achieve high speaker similarity and speech quality, we build the system upon
YourTTS \cite{casanova2022yourtts} and incorporate several modifications from VITS2 \cite{kong23_interspeech}.
The speaker-aware text encoder, flow-based decoder with Transformer blocks, and noise-injected monotonic alignment search are introduced.
For data preprocessing, we resample, normalize and denoise the audios.
For model training, we mix up few-shot data with pre-training data and adopt a speaker-balanced sampling strategy during fine-tuning.
According to the official evaluations, our system achieves the best speaker similarity MOS of 4.25 and obtains considerable naturalness MOS of 3.97 in track 1.

%% file: Paper_Parts/Methodology.tex
\section{Methodology}
\vspace{-0.1cm}
\label{sec:format}
In this part, we will describe our multi-speaker multi-lingual TTS voice cloning system in detail.
\vspace{-0.1cm}

\subsection{Data Preprocessing}
\vspace{-0.1cm}

The competition provides two datasets, \textit{TTS training data} and \textit{few-shot data}, respectively.
The \textit{TTS training data} contains 560 hours of studio-quality TTS data in 7 Indian languages for base model pre-training.
Each language consists of a male and a female speaker, and each speaker has about 40 hours of corpus.
In text processing, we just use the raw character sequence as input, without any text normalization or grapheme-to-phoneme conversion.
After traversing the entire corpus, we get a character set of size 467.
In audio processing, we use the samples with duration between 2 and 15 seconds.
Then we re-sample the waveforms to 16 kHz and normalize the volume. 
We also compute linear-spectrograms from waveforms to guide the training of the posterior encoder.

The \textit{few-shot data} consists of 9 target speakers and each speaker has about 5 minutes speech.
Due to the presence of some environmental noises in part of  them, we conduct speech enhancement for this dataset by FullSubNet+ \cite{chen2022fullsubnet+}.
Afterward, we do the same re-sampling, normalization and computation operations as above.

\vspace{-0.3cm}

\subsection{Model Architecture}
\begin{figure}[htbp]
\centering  \includegraphics[width=0.95\linewidth]{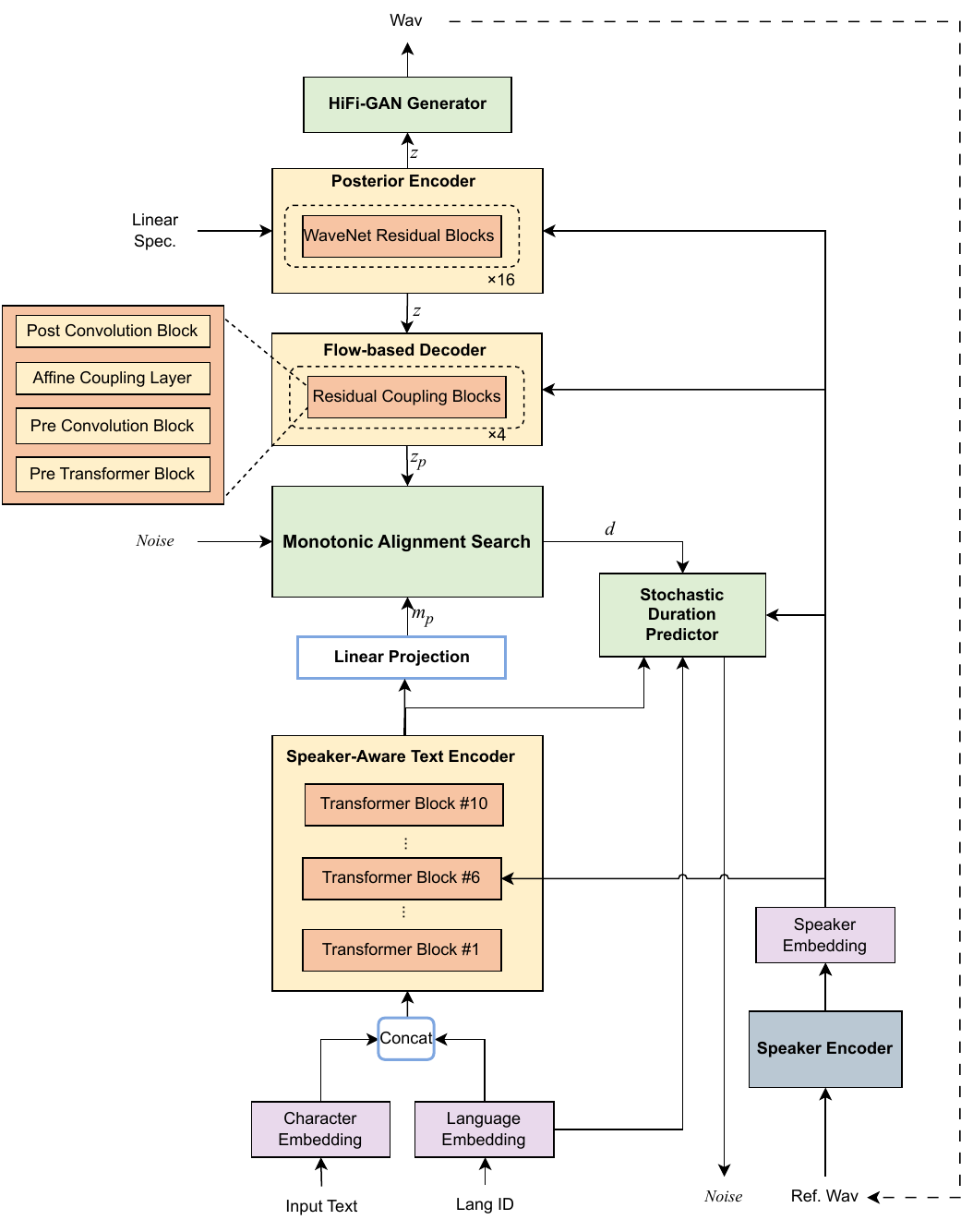}
    \vspace{-0.3cm}
	\caption{The model architecture in training procedure.}
	\label{fig:model}
	\vspace{-0.3cm}
\end{figure}

\vspace{-0.15cm}

As shown in Fig. \ref{fig:model}, the main framework of our system builds upon YourTTS, an end-to-end zero-shot multi-speaker TTS model that can also be used for few-shot scenarios.
We add several modifications inspired by VITS2, to further enhance the performance on speaker similarity and speech quality.

Firstly, the language embeddings are concatenated to each character embedding for multilingual training, and passed to the text encoder.
Here we utilize the speaker-aware text encoder instead of the original one.
Specifically, it incorporates the speaker embedding to the 6-th Transformer block of the text encoder.
It is found that considering speaker embedding when encoding the input text can make model better capture a person's speaking style or particular pronunciation characteristics (i.e., accent).
This helps significantly improve the speaker similarity from the sense of listening.
Next, as in YourTTS, the posterior encoder receives a linear spectrogram and outputs a latent variable $z$ which is used to train the HiFi-GAN generator and the flow-based decoder.
To enable the ability of modelling long-term dependencies, the Transformer block is introduced into the flow-based decoder, that is inserted with residual connections before the convolution block.
The monotonic alignment search (MAS) is employed to learn the alignment between text and speech, and guides the training of the stochastic duration predictor.
In MAS, a gradually decaying Gaussian noise is added when calculating probabilities, for encouraging the model to search other possible alignments.
Due to the limited amount of data of the target speaker, we use a pre-trained speaker encoder to extract speaker embeddings from waveforms, which makes model easier to converge compared to using the learnable speaker embedding table\footnote{The model implementation and the pre-trained speaker encoder are both based on the official baseline: \url{https://github.com/bloodraven66/LIMMITS-24-Coquiai/}}.

In the inference procedure, MAS and the posterior encoder are not used.
We use a randomly selected sample from the target speaker as the reference waveform to provide the speaker embedding.
The duration is predicted by the stochastic duration predictor and the latent variable $z$ is obtained from the inverted flow-based decoder, consistent with YourTTS \cite{casanova2022yourtts}.

\vspace{-0.2cm}
\subsection{Model Training}
\vspace{-0.1cm}
\subsubsection{Pre-training}
\vspace{-0.1cm}
We use 6 NVIDIA 2080 Ti GPUs to train the base model and the batch size is 12 on per GPU.
The base model is trained with the 14 speakers \textit{TTS training data} for 350K iterations.
The initial learning rate is set to 2e-4 with decaying exponentially by a gamma of 0.999875.

\vspace{-0.2cm}
\subsubsection{Few-shot fine-tuning}
\vspace{-0.1cm}
To achieve few-shot voice cloning on target speakers, we fine-tune the base model with the \textit{few-shot data}.
As the model has a large amount of parameters to update, it is easier to overfit the limited data, resulting in great deterioration of the speech intelligibility.
Thus we mix up the \textit{TTS training data} and \textit{few-shot data} during fine-tuning stage \cite{morioka2022residual}.
To guarantee effective training for target speakers, a speaker-balanced sampling strategy is adopted, 
that is, the probability of each speaker is equal when organizing the batch from samples.
In this way, we can fine-tune the whole model without naturalness or quality deterioration.
We use a single GPU to fine-tune the model and the learning rate is set to 1e-4.
After 60K iterations fine-tuning, the model can synthesize speeches with 9 target speakers' voice in both mono-lingual and cross-lingual scenarios.

%% file: Paper_Parts/Results.tex
\vspace{-0.1cm}
\section{Results}
\vspace{-0.1cm}

Since LIMMITS'24 Challenge involves voice cloning, the evaluation includes naturalness and speaker similarity mean opinion scores (MOS).
The test set utterances consist of both mono-lingual and cross-lingual synthesis, covering all 7 Indian languages.
We only participate in track 1 and the MOS scores in the subjective listening tests are shown in table \ref{table:1}.
In track 1, our system achieves the best speaker similarity result, significantly outperforming the other teams with the gap of more than 0.2, and obtains the respectable naturalness result.
Besides, our speaker similarity MOS is also superior than all teams in track 2 which could use external datasets to perform few-shot voice cloning.

\vspace{-0.2cm}

\begin{table}[!htbp]
    \vspace{-0.1cm}
    \begin{center}
    \caption{MOS results in LIMMITS'24 Challenge track 1.
    } 
    \label{table:1}
    \scalebox{0.9}{
    \begin{tabular}{ccc}
    \toprule
        & Score (avg) & Score (std) \\
    \bottomrule
    Naturalness &  3.97 & 0.92 \\
    Speaker Similarity &  4.25 & 1.06   \\
    \bottomrule
    \end{tabular}}
    \begin{tablenotes}
     \ninept{\item[1] $*$ More detailed results are available on this \href{https://sites.google.com/view/limmits24/results}{website}.}
   \end{tablenotes}
    \end{center}
    \vspace{-0.05cm}
\end{table}

\vspace{-0.3cm}

%% file: Paper_Parts/Conclusions.tex
\vspace{-0.2cm}
\section{Conclusions}
\vspace{-0.2cm}
\label{sec:conclusions}
This paper introduces the THU-HCSI multi-speaker multi-lingual few-shot voice cloning system for LIMMITS'24 Challenge. 
Our system is built upon YourTTS with several useful modifications from VITS2.
We also adopt some effective data preprocessing operations and fine-tuning strategies to improve the speech quality.
The proposed system is ranked 1st in terms of speaker similarity in track 1, and obtains considerable naturalness results.